\documentclass[aps,prl,twocolumn,superscriptaddress,longbibliography]{revtex4-2}
\usepackage{amsmath,amssymb}
\usepackage[pdftex]{hyperref,graphicx}
\hypersetup{colorlinks = true, urlcolor = blue, linkcolor = blue, citecolor = blue}
\usepackage{physics}
\usepackage{xcolor}
\usepackage{bm}
\usepackage{pdfpages}

\newcommand{\comment}[1]{{}}

\makeatletter
\AtBeginDocument{\let\LS@rot\@undefined}
\makeatother

\begin{document}
\title{Valley polarization transition in a two-dimensional electron gas}
\author{Seongjin Ahn}
\affiliation{Condensed Matter Theory Center and Joint Quantum Institute, Department of Physics, University of Maryland, College Park, Maryland 20742, USA}
\author{Sankar Das Sarma}
\affiliation{Condensed Matter Theory Center and Joint Quantum Institute, Department of Physics, University of Maryland, College Park, Maryland 20742, USA}

\begin{abstract}
We theoretically study transport signatures associated with a spontaneous 2-valley to 1-valley quantum phase transition in a two-dimensional electron gas (2DEG) tuned by decreasing the 2D carrier density, as claimed in a recent experiment [\href{https://doi.org/10.1103/PhysRevLett.127.116601}{Phys. Rev. Lett. 127, 116601 (2021)}]. The key issue we focus on is whether the experimentally measured 2D resistivity as a function of carrier density is consistent (or not) with an underlying spontaneous valley-polarization transition as assumed uncritically in the experimental report.  Our theoretical analysis is particularly germane since the experiment does not directly measure the change in the Fermi surface resulting from the valley polarization transition, but infers such a transition indirectly through transport measurements. We validate the experimental claim, showing that indeed the observed sudden change in the 2D resistivity is quantitatively consistent with a sudden change in the valley polarization from 2 to 1 at the critical density.

\end{abstract}

\maketitle
{\em Introduction.}--- 
The possibility of a density-driven spin-polarization transition in an electron gas from a paramagnetic spin-unpolarized (i.e., spin degeneracy 2) to a ferromagnetic spin-polarized (i.e., spin-degeneracy 1) system with the decreasing of density was predicted by Bloch almost 100 years ago, and is historically the first quantum many body prediction \cite{blochBemerkungZurElektronentheorie1929}. The basic idea is that at high density the electrons occupy both spin states equally so as to minimize the kinetic energy whereas at low density, electrons occupy only one spin state so as to minimize the exchange energy \cite{vuReentrantBlochFerromagnetism2021}. 
Although Bloch’s original hope of explaining the metallic ferromagnetism of Fe, Ni, Co has turned out to be incorrect since it is now known that the narrow band nature of the transition metals plays a crucial role in metallic ferromagnetism, the fact that exchange (kinetic) energy enhances ferromagnetism (paramagnetism) by preferring 2-spin (1-spin) electron occupancy is well-established.  Whether a Bloch-type strictly density-driven paramagnetic-to-ferromagnetic spin transition happens or not is still an open question, but any such transition can only occur at extreme low carrier densities which are inaccessible in any 3DEG systems. Recently, such a Bloch spin transition has been experimentally reported in 2D composite fermions at the half Landau level filling \cite{hossainBlochFerromagnetismComposite2021, vuReentrantBlochFerromagnetism2021}. 
 
Semiconductors often have additional degrees of freedom beyond spin, such as valleys, arising from the detailed nature of their conduction (or valence) bands where multiple equivalent degenerate energy minima may exist in their band structures.  Well-known examples of such valley degeneracy are Si, Ge, AlAs conduction bands.  The corresponding 2DEG in these materials will then have a certain known valley degeneracy in the noninteracting situation.  For example, Si 100, 110, and 111 surface 2DEG systems have valley degeneracies of 2, 4, and 6 respectively \cite{andoElectronicPropertiesTwodimensional1982}.  
In such valley-degenerate systems, the valley symmetry may be spontaneously broken by many body effects as the density is lowered similar to the `Bloch ferromagnetism' except that the density-driven transition is now a `valley-polarization' transition with the system spontaneously transitioning from a high-density high-valley-degeneracy ground state to a low-valley-degeneracy low-density ground state in order to minimize the exchange energy. Such valley-polarization transitions were envisioned and predicted in Si-based 2DEG a long time ago \cite{dassarmaManybodyCorrelationEffects1983, blossInteractionInducedTransitionLow1979},
and there were even experimental claims of observing such a valley polarization transition \cite{coleEvidenceValleyOccupancyTransition1981} 
although the situation remained unclear because Si 2D systems often have intrinsic surface-induced valley splitting at all densities which have nothing to do with the exchange-driven density-tuned valley polarization transition \cite{andoElectronicPropertiesTwodimensional1982, tsuiObservationSixfoldValley1979}. 
 
In this context, it is therefore highly interesting that a recent experimental publication claims the observation of the elusive density-tuned spontaneous itinerant electron valley-polarization transition in the AlAs-based high-mobility 2DEG \cite{hossainSpontaneousValleyPolarization2021}. 
If validated, this is an important claim since it would presumably be the first observation of `Bloch ferromagnetism' albeit in the form of `valley-magnetism'. The current work is focused on validating (or invalidating) the experimental claim by carefully analyzing the experimental transport data below and above the experimentally claimed critical carrier density in order to check whether the theoretically calculated density-dependent resistivity is indeed consistent with a spontaneous valley polarization transition. This is necessary since the experiment does not measure either the valley degeneracy or the Fermi surface size (i.e., Fermi momentum) directly, and the experimental conclusion on a spontaneous valley polarization transition is based entirely on the indirect evidence of a sudden change in the 2D resistivity at a critical density. Our transport theory can, in principle, validate whether the experimental resistivity is consistent with a sudden change in the valley degeneracy of the 2D AlAs system. 
 
Such a theoretical validation takes on particular significance in view of the fact that there are contradictory theoretical predictions, based on ground state energetic calculations, not only for the critical density for such a transition \cite{zhangDensitydependentSpinSusceptibility2005, attaccaliteCorrelationEnergySpin2002, zhangCommentEffectsThickness2006},
but whether such a transition even takes place at all at any density \cite{tanatarGroundStateTwodimensional1989,drummondPhaseDiagramLowDensity2009}. 
It is therefore of considerable importance to critically validate the experimental claim of spontaneous valley polarization in a 2DEG because the state of the art quantum Monte Carlo calculations conclude that such a Bloch ferromagnetic transition should not occur with a categorical statement right in the abstract saying `the fully spin-polarized fluid is never stable' \cite{drummondPhaseDiagramLowDensity2009}!  
The nominal observation of the 2-valley to 1-valley spontaneous valley polarization transition directly challenges this theoretical assertion. So, our theoretical work is by no means routine, if we validate the experimental claim (as we in fact do in the rest of this paper), new thinking would be necessary about the low-density physics of 2DEG, and the meaning of a many body valley polarization transition would have to thoroughly reconsidered.  We emphasize that our work is purely phenomenological as we accept the experimental claim of a spontaneous valley polarization transition, and then calculate the density-dependent system resistivity, comparing our theoretical results directly with the experimental data in order to conclude about the correctness of the experimental claim. Our work thus avoids the difficult challenge of calculating the ground state energies for competing states (which are often extremely close in energy) in order to predict the actual ground state precisely. Our goal is to validate or invalidate the experimental claim of the system being in a 2-valley (1-valley) state above (below) the critical density based only on transport considerations.

\begin{figure*}[!htb]
\centering
\includegraphics[width=1.0\linewidth]{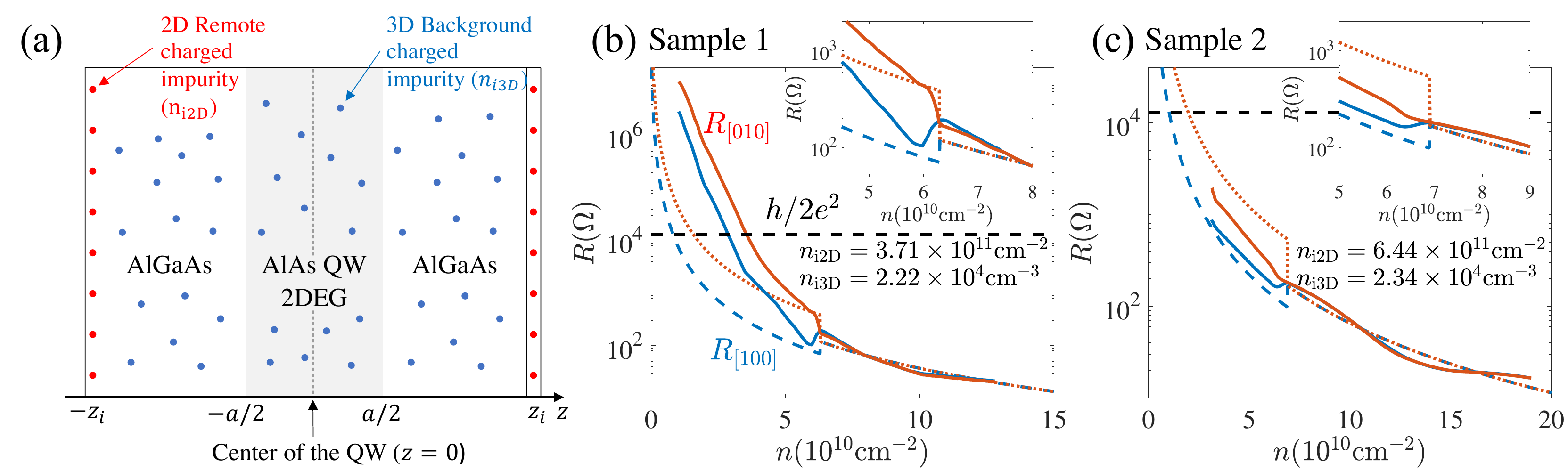}
\caption{(a) Schematic of the sample structure. $a=21\mathrm{nm}$ ($20\mathrm{nm}$) is the width of the 2DEG AlAs quantum well(QW) for Sample 1 (2). Blue circles represent charged background impurities randomly distributed throughout the sample from $z=-z_i$ to $z=z_i$, where $z_i=a/2 + 68\mathrm{nm}$ with $68\mathrm{nm}$ being the thickness of the AlGaAs spacer layer. Red circles represent remote charged impurities in the 2D layer located at $z=\pm z_i$. (b), (c) Experimental resistivity (solid lines) for (b) Sample 1 and (c) Sample 2 along the [010] (red) and [100] (blue) directions plotted as a function of the carrier density $n$ along with the best theoretical fit (dashed lines) obtained using the Boltzmann theory with two tuning parameters ($n_\mathrm{i2D}$, $n_\mathrm{i3D}$). The inset presents a zoom-in around the critical density where the valley transition occurs. The best fit parameter values are shown in the figures.
  }
\label{fig:1}
\end{figure*}

{\em Theory and Results.}---  
We use the highly successful Boltzmann transport theory for the resistivity calculations at $T=0$ since the experiments are carried out at 20 mK. The only resistive scattering mechanism operational at such low temperatures is the disorder or impurity scattering. It is well-known that such high-mobility modulation-doped 2D semiconductor structures typically have two distinct types of disorder scattering, both associated with random quenched charged impurity centers, the so-called `background impurity' and `remote impurity' \cite{dassarmaTransportTwodimensionalModulationdoped2015}. 
The background disorder arises from unintentional impurities in the system distributed three-dimensionally throughout the structure whereas the remote disorder is associated with the 2D modulation doping layer deliberately put in the system at a distance from the 2DEG so as to minimize the scattering between the charged dopants and the electrons, allowing the achievement of very high mobility in the system.  Since the actual AlAs structure used in the experiment is known, we can carry out a 2-parameter realistic calculation, where the two unknown parameters are the 3D background impurity density $n_\mathrm{i3d}$ and the 2D remote impurity density $n_\mathrm{i2d}$ placed at a fixed distance from the quantum well as in the experimental sample [see Fig.~\ref{fig:1}(a) for a schematic of the system]. We then carry out a realistic resistivity calculation where the resistivity `$R$' is given by the Drude formula 
\begin{equation} \label{eq:Drude_Formula}
R = \frac{m}{ne^2\tau}
\end{equation}
where $m$ is the effective mass and $1/\tau$ is the transport scattering rate. Since the experimental resistivity is anisotropic below the critical density where only a single valley with anisotropic mass is occupied [see Figs.~\ref{fig:1}(b) and ~\ref{fig:1}(c)], we set $m=0.2$($1.1$) in Eq.~(\ref{eq:Drude_Formula}) for calculating the resistivity along the [100]([010]) direction, following the experimental paper \cite{hossainBlochFerromagnetismComposite2021}. Above the critical density the experimental resistivity is isotropic and thus we use the isotropic averaged density of states mass defined as $m_\mathrm{DOS}=\sqrt{0.2\times1.1}=0.46$.
For the calculation of $1/\tau$, we make the isotropic approximation (even at low densities where mass anisotropy is present), meaning that we use the isotropic Boltzmann theory with the anisotropic effective mass approximated as the isotropic averaged mass $m_\mathrm{DOS}$ instead of solving the full anisotropic Boltzmann transport equation. This isotropic approach is well justified given that the system is in the very low density regime with $k_\mathrm{F}/q_\mathrm{TF}\sim 0.01\ll1$, where the isotropic approximation is valid \cite{ahnScreeningFriedelOscillations2021}. Here $q_\mathrm{TF}$ is the Thomas-Fermi screening wave vector. The transport scattering rate within the Born approximation is given by
\begin{equation} \label{eq:transport_relaxation_time}
    \frac{1}{\tau}=\frac{2\pi}{\hbar} \int N_i(z) dz 
    \sum_{\bm k'}
    \left|V_{\bm k - \bm k'}(z)\right|^2
    (1-\cos{ \theta})\delta(\epsilon_{\bm k_\mathrm{F}}-\epsilon_{\bm k'}),
\end{equation}
where $\epsilon_{\bm k}=\hbar^2k^2/2m_{\mathrm{DOS}}$, $\bm k_\mathrm{F}$ is the Fermi wave vector, $N_i(z)$ is the distribution of impurities, $z$ is the distance from the center of the quantum well [see Fig.~\ref{fig:1}(a)], $\theta$ is the scattering angle between $\bm k$ and $\bm k'$, and $V_{\bm q}(z)=(v^{(c)}_{\bm q}/\varepsilon_{\bm q}) e^{-q\left|z\right|}F^{(i)}_{\bm q}$ is the electron-impurity scattering matrix element and $v^{(c)}_{\bm q}=2\pi e^2/\kappa q$ is the Coulomb interaction with $\kappa$ being the dielectric constant. $\varepsilon_{\bm q}=1+v^{(c)}_{\bm q} F_{\bm q} \Pi_{\bm q}$ is the screening function, where $\Pi_{\bm q}$ is the noninteracting polarzaibility given by \cite{sternPolarizabilityTwoDimensionalElectron1967}
 \begin{equation}
    \Pi_{\bm q}=
		\frac{gm_\mathrm{DOS}}{2\pi\hbar^2}\left[1 - \Theta(q-2k_\mathrm{F})\frac{\sqrt{q^2- 4k^2_\mathrm{F} }}{q}\right],
    \label{eq:diel}
\end{equation}
and $g$ is the degeneracy factor.
For realistic calculations, we incorporate the effects of quantum well thickness through the form factors $F^{(i)}_{\bm q}$ and $F_{\bm q}$ given by \cite{hwangElectronicTransportTwodimensional2013}
\begin{equation}
    F_{\bm q}=\frac{ 3(qa) + 8\pi^2/(qa)}{(qa)^2 + 4\pi^2} - \frac{32\pi^4(1-e^{-qa})}{(qa)^2[(qa)^2 + 4\pi^2 ]^2},
\end{equation}
and
\begin{equation} \label{eq:5}
    F^{(i)}_{\bm q}= \frac{4}{qa} \frac{ 2\pi^2(1-e^{-qa/2}) + (qa)^2 }{ (4\pi)^2 + (qa)^2 },
\end{equation}
where $a$ is the quantum well width. 




\begin{figure}[!htb]
\centering
\includegraphics[width=1.0\linewidth]{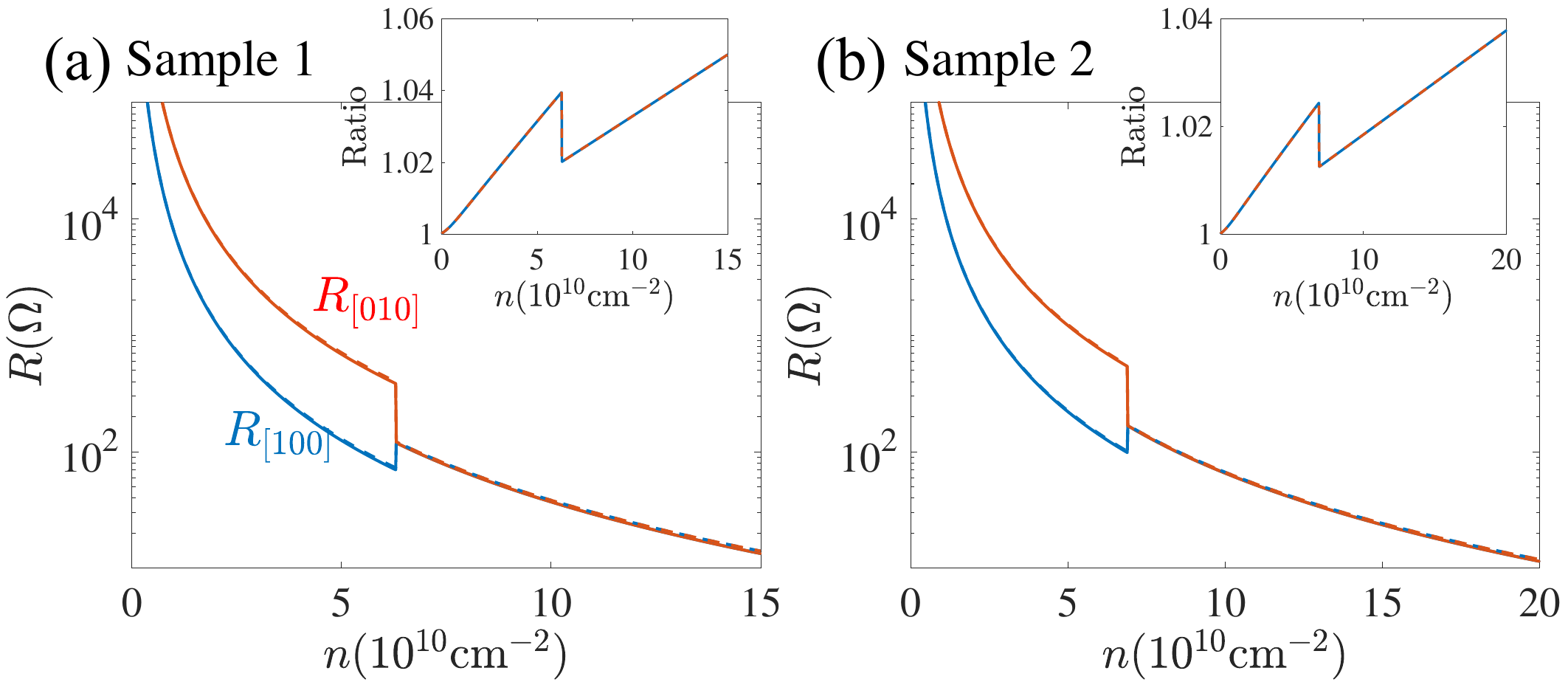}
\caption{The best-fit theoretical resistivities obtained in Fig.~\ref{fig:1} (solid line) in comparison with those computed using the same best-fit remote impurity density ($n_\mathrm{i2D}$) but with a larger background impurity density of $n_\mathrm{i3D}=10^{13}\mathrm{cm}^{-3}$ (dashed line) for (a) Sample 1 and (b) Sample 2. The inset shows the ratio of the resistivity calculated using $n_\mathrm{i3D}=10^{13}\mathrm{cm}^{-3}$ to the best-fit resistivity with the dashed line in red (blue) color indicating the direction [010] ([100]).
  }
\label{fig:2}
\end{figure}

\begin{figure*}[!htb]
\centering
\includegraphics[width=1.0\linewidth]{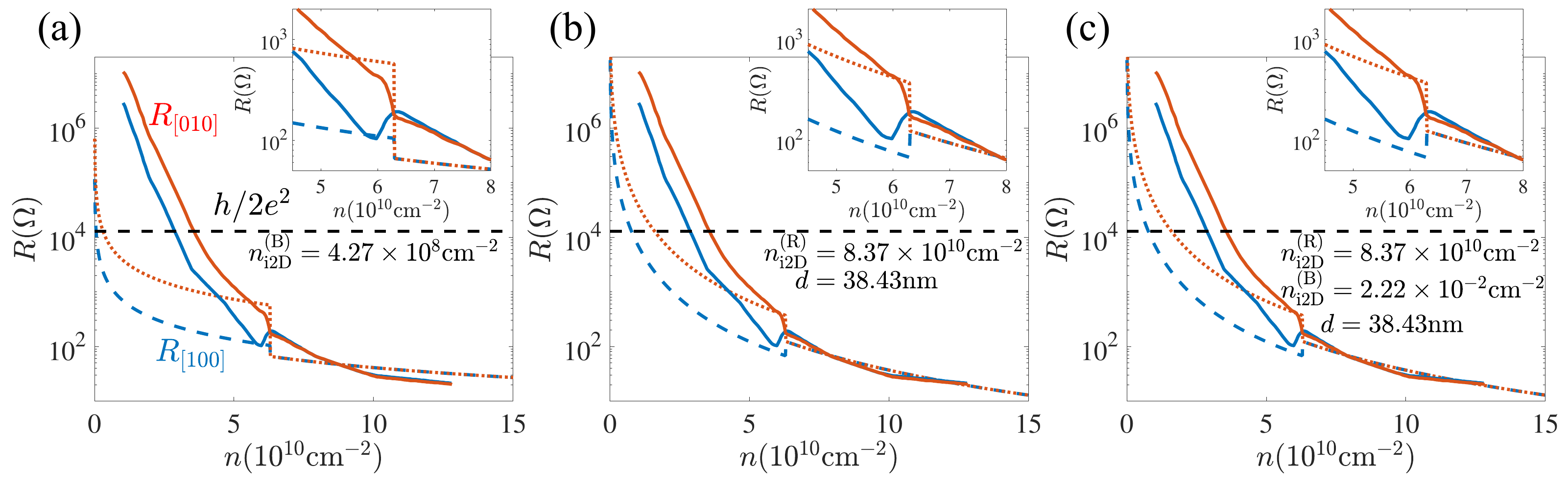}
\caption{Experimental resistivity and the best theoretical fits for Sample 1 calculated in the strict 2D limit. The notations are the same as in Fig.~\ref{fig:1}. The theoretical results are obtained assuming scattering mechanisms associated with (a) background charged impurities of density $n^\mathrm{(B)}_\mathrm{i2D}$ distributed in the 2DEG layer, (b) remote charged impurities of density $n^\mathrm{(R)}_\mathrm{i2D}$ located at a distance of $d$ away from the 2DEG layer, and (c) both background and remote charged impurities considered in (a) and (b). 
  }
\label{fig:3}
\end{figure*}

The calculated resistivity depends on $n, n_\mathrm{i3d}$ and $n_\mathrm{i2d}$ [i.e., $R=R$($n, n_\mathrm{i3d}, n_\mathrm{i2d}$)], where $n$ is the variable 2D electron density, which is the experimental tuning parameter controlling the 2D resistivity, and $n_\mathrm{i3d}$ and $n_\mathrm{i2d}$ are the unknown Coulomb impurity densities, which we vary to obtain the best fits to the experimental resistivity as shown in Fig.~\ref{fig:1}. Our recursive fitting procedure to fix the disorder parameters demands as good a density-dependent resistivity fit to the experimental data as possible at the highest experimental density where our Boltzmann theory is perfectly valid because the $k_\mathrm{F}l\gg1$ approximation (where $l$ is the transport mean free path) applies. Once the best fit is done for the high-density data (note that we reproduce this high-density 2-valley degeneracy experimental results perfectly), we no longer adjust $n_\mathrm{i3d}$ and $n_\mathrm{i2d}$ and simply carry out the calculations to lower densities with no further adjustments of $n_\mathrm{i3d}$ and $n_\mathrm{i2d}$. The only adjustments we do are, following the experimental claim, to change the valley degeneracy from 2 to 1 and the effective mass in the Drude formula [i.e., $m$ in Eq.~(\ref{eq:Drude_Formula})] from the in-plane isotropic effective mass of $0.46$ to $0.2$ ($1.1$) for the [100] ([010]) direction at the critical density $n=n_\mathrm{c}$ in our calculations. 
 
In Figs.~\ref{fig:1}(b) and \ref{fig:1}(c), we present the unbiased fitting of our theoretical results to the measured experimental resistivity in both experimental samples (Samples 1 and 2) presented in Ref.~\cite{hossainSpontaneousValleyPolarization2021}, finding the best fit impurity densities to be; $n_\mathrm{i2d}= 3.71\times10^{11}\mathrm{cm}^{-2}$ $(6.44\times10^{11}\mathrm{cm}^{-2}) $ and $n_\mathrm{i3d} =2.22\times10^4 \mathrm{cm}^{-3}$ $(2.34\times10^4\mathrm{cm}^{-3})$  in Samples 1 (2). Note that the unbiased best fit value for the background unintentional impurity density is basically zero, which is consistent with the known extreme high quality of the AlAs 2DEG system used in the experiment.  To check the robustness of this fit, we show in Fig.~\ref{fig:2} the fit to the experimental resistivity by changing $n_\mathrm{i3d}$ to $10^{13} \mathrm{cm}^{-3}$, i.e., increasing the background disorder by 15 orders of magnitude. The theoretical results hardly change, clearly establishing that the resistivity of the 2DEG is determined entirely by the remote charged dopant scattering.
 
Since the results presented in Figs.~\ref{fig:1} and \ref{fig:2} are our central findings, we now discuss their salient features: (1) Experiment and theory agree exactly at high densities ($n>n_\mathrm{c}$), and the agreement is semi-quantitative at lower densities, which is understandable since the dimensionless $k_\mathrm{F}l$ parameter becomes progressively smaller with decreasing carrier density making the Boltzmann theory less applicable in the low-density $n<n_\mathrm{c}$ regime—we emphasize that our fit uses the high-density regime where the theory is exact. (2) The most important aspect of our results in the current context is that the theory well reproduces the experimentally observed resistive discontinuity at the critical density with an abrupt jump or drop in the resistivity at $n=n_\mathrm{c}$ depending on the direction of transport (because of the mass anisotropy)—the importance of this finding is that the theory has the valley polarization transition at $n=n_\mathrm{c}$ explicitly included in the calculations, thus manifestly validating the experimental claim of a spontaneous valley polarization transition at $n=n_\mathrm{c}$. (3) The theoretical resistivity discontinuity at $n=n_\mathrm{c}$ is in reasonable quantitative agreement with the experiment with the experimental discontinuity being slightly smaller and somewhat smoother than the sharp theoretical discontinuity, most likely because of thermal broadening. (4) The theory underestimates the measured resistivity with decreasing density for $n<n_\mathrm{c}$, but the resistivity at the critical density is still way below the Mott-Ioffe-Regel limit and satisfies well the Boltzmann theory validity criterion of $k_\mathrm{F}l>1$.  (5) Overall the theory and experiment are in remarkable good agreement providing a strong validation of the experimental claim of the manifestation of a density-tuned spontaneous 2D valley polarization transition.
 
To further reinforce the agreement between theory and experiment validating the 2D valley polarization transition, we show in Fig.~\ref{fig:3} three different resistivity calculation results for Sample 1, all carried out in the strict 2D limit, where the 2DEG is assumed to be a simple zero-thickness 2D layer of electrons and the disorder is treated in three different models:  (1) 1-parameter model, where the impurities are located in the 2DEG layer and the only free parameter is the 2D random charged impurity density; (2)  2-parameter model, where all the random impurities are located in a 2D impurity layer at a distance of `$d$' from the 2DEG, and the two parameters are the 2D impurity density and `$d$'; (3) 3-parameter model, which combines the 1- and 2-parameter models, with two different impurity densities located in the 2DEG and a distance `$d$' away. The results in Fig.~\ref{fig:3} lead decisively to two distinct conclusions: (1) clearly the best disorder model is the 2-parameter model, where the impurities are located at a layer away from the 2DEG, which is of course completely consistent with our realistic results presented in Figs.~\ref{fig:1} and \ref{fig:2}, showing that the resistive scattering in the experiment is completely dominated by remote dopant scattering; (2) the manifestation of the spontaneous valley polarization transition in the measured resistivity at the critical density is independent of the disorder model used in the theory—all three simple models used in Fig.~\ref{fig:3} as well as the realistic models used for Figs.~\ref{fig:1} and \ref{fig:2} show agreement with the experiment, manifestly validating the experimental claim of a density-tuned 2-valley to 1-valley transition in the AlAs 2DEG system.
 
Before concluding, we mention several features of the theory which requires emphasis. First, our use of random charged impurity scattering as the disorder mechanism is completely consistent with the known scattering mechanisms in MBE-grown high-mobility semiconductor structures in general and AlAs quantum wells in particular. We have redone our theoretical analyses adding a short-range disorder term, and this short-range disorder strength turns out to be zero when theory and experiment are compared, which is completely understandable because of the extreme high quality of the AlAs quantum well system employed in the experiment. In addition, short-range disorder leads to a resistivity varying strictly as inverse carrier density for all densities (because the scattering rate is independent of the density), which disagrees with the experimental data. Even more importantly, short range disorder would lead to a complete valley degeneracy independent resistivity with NO manifestation of the valley polarization transition in the resistivity (except through the effective mass anisotropy).  Second, the fact that the experimental sample is dominated by remote impurity scattering is apparent from the resistivity behavior at the transition, where the resistivity $R_\mathrm{[100]}$ along the 100 direction actually drops at the transition (in contrast to the $R_\mathrm{[010]}$). This drop is inconsistent with background impurity scattering, and is only consistent with remote impurity scattering. In particular, it is straightforward to show analytically, using Eqs.~(\ref{eq:Drude_Formula})-(\ref{eq:5}), that the resistivity goes asymptotically as $m/g^2$ in the strong screening limit.
At the putative phase transition, the effective mass changes from $0.46$ to $0.2$ for [100] direction transport, but the valley degeneracy decreases by a factor of 2, which would then lead to a factor of $1.7$ increase in $R_\mathrm{[100]}$ as can be seen in Fig.~\ref{fig:3}(a).  Experimentally, however, $R_\mathrm{[100]}$ drops at the critical density, which cannot therefore be explained by the background impurity scattering.  For remote doping, however, the change in valley degeneracy also affects the factor $k_\mathrm{F}d$ in the exponential term in the integral since the Fermi wavenumber depends explicitly on the valley degeneracy, $k_\mathrm{F}\sim g^{-1/2}$, $k_\mathrm{F}d$ thus becoming larger when the valley degeneracy is reduced, leading to a decrease in the resistivity $R_\mathrm{[100]}$ as observed experimentally. For $R_\mathrm{[010]}$, the effective mass itself increases at the transition, thus leading to an increase in $R_\mathrm{[010]}$ both from the mass and from the valley degeneracy change.  Thus, the experimental behavior of the valley polarization transition itself tells us the type of scattering dominating transport in the system.  It may also be useful to point out that our calculated resistivity for Sample 1 in Figs.~\ref{fig:1} and \ref{fig:2} reaches the Mott-Ioffe-Regel strong localization limit of $k_\mathrm{F}l=1$ at the density of $3\times10^{10} \mathrm{cm}^{-2}$ to be compared with the experimental metal-insulator transition density of $4\times10^{10} \mathrm{cm}^{-2}$ whereas the valley polarization critical density is $6\times10^{10} \mathrm{cm}^{-2}$.  For Sample 2, we predict a metal-insulator transition critical density of $2.5\times10^{10} \mathrm{cm}^{-2}$, which also appears to be close to the experimental finding whereas the valley polarization critical density is $7\times10^{10} \mathrm{cm}^{-2}$. Our theory also implies that the observed low-density transition to the insulator is an Anderson-type strong localization crossover.

{\em Conclusion.}---  
We have critically analyzed the transport experiment of Ref.~\cite{hossainSpontaneousValleyPolarization2021} which reports the density-tuned 2D spontaneous valley polarization transition through the measurement of low-temperature density-dependent 2D resistivity. We validate the experimental claim quantitatively, establishing that, although the experiment did not directly measure either the ground state valley degeneracy or the Fermi surface shape, the reported resistivity data indeed correlate directly and quantitatively with a sudden change in the valley occupancy from 2 to 1 at the critical density. This validation takes on particular significance because the best quantum Monte Carlo calculations do not predict such a valley polarization transition (and the ones that do predict it to happen at lower densities than the experimentally reported critical density). So, although the observation is definitely consistent with a density-tuned spontaneous valley polarization transition, the question remains why it is happening at all. The simplest and the most obvious possibility is that the quantum Monte Carlo predictions are perhaps not precisely reliable. Another possibility is that the system already has small valley splitting even in the non-interacting high density situation, which is enhanced by interaction. This possibility cannot be ruled out, particularly since it is well-known that in Si-based 2D systems, the valley degeneracy is always slightly lifted by single-particle band structure effects, and small random valley splittings are always present in every sample, depending on the details of the interface, disorder, and carrier density \cite{andoElectronicPropertiesTwodimensional1982}. More experiments would be necessary to resolve this issue, but the fact that Ref.~\cite{hossainSpontaneousValleyPolarization2021} observed a density-tuned valley polarization transition is unquestionable.

\section{Acknowledgement} \label{sec:acknowledgement}
This work is supported by the Laboratory for Physical Sciences.

\bibliography{ref}

\begin{thebibliography}{18}%
\makeatletter
\providecommand \@ifxundefined [1]{%
 \@ifx{#1\undefined}
}%
\providecommand \@ifnum [1]{%
 \ifnum #1\expandafter \@firstoftwo
 \else \expandafter \@secondoftwo
 \fi
}%
\providecommand \@ifx [1]{%
 \ifx #1\expandafter \@firstoftwo
 \else \expandafter \@secondoftwo
 \fi
}%
\providecommand \natexlab [1]{#1}%
\providecommand \enquote  [1]{``#1''}%
\providecommand \bibnamefont  [1]{#1}%
\providecommand \bibfnamefont [1]{#1}%
\providecommand \citenamefont [1]{#1}%
\providecommand \href@noop [0]{\@secondoftwo}%
\providecommand \href [0]{\begingroup \@sanitize@url \@href}%
\providecommand \@href[1]{\@@startlink{#1}\@@href}%
\providecommand \@@href[1]{\endgroup#1\@@endlink}%
\providecommand \@sanitize@url [0]{\catcode `\\12\catcode `\$12\catcode
  `\&12\catcode `\#12\catcode `\^12\catcode `\_12\catcode `\%12\relax}%
\providecommand \@@startlink[1]{}%
\providecommand \@@endlink[0]{}%
\providecommand \url  [0]{\begingroup\@sanitize@url \@url }%
\providecommand \@url [1]{\endgroup\@href {#1}{\urlprefix }}%
\providecommand \urlprefix  [0]{URL }%
\providecommand \Eprint [0]{\href }%
\providecommand \doibase [0]{https://doi.org/}%
\providecommand \selectlanguage [0]{\@gobble}%
\providecommand \bibinfo  [0]{\@secondoftwo}%
\providecommand \bibfield  [0]{\@secondoftwo}%
\providecommand \translation [1]{[#1]}%
\providecommand \BibitemOpen [0]{}%
\providecommand \bibitemStop [0]{}%
\providecommand \bibitemNoStop [0]{.\EOS\space}%
\providecommand \EOS [0]{\spacefactor3000\relax}%
\providecommand \BibitemShut  [1]{\csname bibitem#1\endcsname}%
\let\auto@bib@innerbib\@empty
\bibitem [{\citenamefont
  {Bloch}(1929)}]{blochBemerkungZurElektronentheorie1929}%
  \BibitemOpen
  \bibfield  {author} {\bibinfo {author} {\bibfnamefont {F.}~\bibnamefont
  {Bloch}},\ }\bibfield  {title} {\bibinfo {title} {{Bemerkung zur
  Elektronentheorie des Ferromagnetismus und der elektrischen
  Leitfähigkeit}},\ }\href {https://doi.org/10.1007/BF01340281} {\bibfield
  {journal} {\bibinfo  {journal} {Z. Physik}\ }\textbf {\bibinfo {volume}
  {57}},\ \bibinfo {pages} {545} (\bibinfo {year} {1929})}\BibitemShut
  {NoStop}%
\bibitem [{\citenamefont {Vu}\ and\ \citenamefont
  {Das~Sarma}(2021)}]{vuReentrantBlochFerromagnetism2021}%
  \BibitemOpen
  \bibfield  {author} {\bibinfo {author} {\bibfnamefont {D.}~\bibnamefont
  {Vu}}\ and\ \bibinfo {author} {\bibfnamefont {S.}~\bibnamefont {Das~Sarma}},\
  }\bibfield  {title} {\bibinfo {title} {Reentrant {{Bloch}} ferromagnetism},\
  }\href {https://doi.org/10.1103/PhysRevB.104.L100405} {\bibfield  {journal}
  {\bibinfo  {journal} {Phys. Rev. B}\ }\textbf {\bibinfo {volume} {104}},\
  \bibinfo {pages} {L100405} (\bibinfo {year} {2021})}\BibitemShut {NoStop}%
\bibitem [{\citenamefont {Hossain}\ \emph
  {et~al.}(2021{\natexlab{a}})\citenamefont {Hossain}, \citenamefont {Zhao},
  \citenamefont {Pu}, \citenamefont {Mueed}, \citenamefont {Ma}, \citenamefont
  {Villegas~Rosales}, \citenamefont {Chung}, \citenamefont {Pfeiffer},
  \citenamefont {West}, \citenamefont {Baldwin}, \citenamefont {Jain},\ and\
  \citenamefont {Shayegan}}]{hossainBlochFerromagnetismComposite2021}%
  \BibitemOpen
  \bibfield  {author} {\bibinfo {author} {\bibfnamefont {M.~S.}\ \bibnamefont
  {Hossain}}, \bibinfo {author} {\bibfnamefont {T.}~\bibnamefont {Zhao}},
  \bibinfo {author} {\bibfnamefont {S.}~\bibnamefont {Pu}}, \bibinfo {author}
  {\bibfnamefont {M.~A.}\ \bibnamefont {Mueed}}, \bibinfo {author}
  {\bibfnamefont {M.~K.}\ \bibnamefont {Ma}}, \bibinfo {author} {\bibfnamefont
  {K.~A.}\ \bibnamefont {Villegas~Rosales}}, \bibinfo {author} {\bibfnamefont
  {Y.~J.}\ \bibnamefont {Chung}}, \bibinfo {author} {\bibfnamefont {L.~N.}\
  \bibnamefont {Pfeiffer}}, \bibinfo {author} {\bibfnamefont {K.~W.}\
  \bibnamefont {West}}, \bibinfo {author} {\bibfnamefont {K.~W.}\ \bibnamefont
  {Baldwin}}, \bibinfo {author} {\bibfnamefont {J.~K.}\ \bibnamefont {Jain}},\
  and\ \bibinfo {author} {\bibfnamefont {M.}~\bibnamefont {Shayegan}},\
  }\bibfield  {title} {\bibinfo {title} {Bloch ferromagnetism of composite
  fermions},\ }\href {https://doi.org/10.1038/s41567-020-1000-z} {\bibfield
  {journal} {\bibinfo  {journal} {Nat. Phys.}\ }\textbf {\bibinfo {volume}
  {17}},\ \bibinfo {pages} {48} (\bibinfo {year}
  {2021}{\natexlab{a}})}\BibitemShut {NoStop}%
\bibitem [{\citenamefont {Ando}\ \emph {et~al.}(1982)\citenamefont {Ando},
  \citenamefont {Fowler},\ and\ \citenamefont
  {Stern}}]{andoElectronicPropertiesTwodimensional1982}%
  \BibitemOpen
  \bibfield  {author} {\bibinfo {author} {\bibfnamefont {T.}~\bibnamefont
  {Ando}}, \bibinfo {author} {\bibfnamefont {A.~B.}\ \bibnamefont {Fowler}},\
  and\ \bibinfo {author} {\bibfnamefont {F.}~\bibnamefont {Stern}},\ }\bibfield
   {title} {\bibinfo {title} {Electronic properties of two-dimensional
  systems},\ }\href {https://doi.org/10.1103/RevModPhys.54.437} {\bibfield
  {journal} {\bibinfo  {journal} {Rev. Mod. Phys.}\ }\textbf {\bibinfo {volume}
  {54}},\ \bibinfo {pages} {437} (\bibinfo {year} {1982})}\BibitemShut
  {NoStop}%
\bibitem [{\citenamefont {Das~Sarma}\ and\ \citenamefont
  {Vinter}(1983)}]{dassarmaManybodyCorrelationEffects1983}%
  \BibitemOpen
  \bibfield  {author} {\bibinfo {author} {\bibfnamefont {S.}~\bibnamefont
  {Das~Sarma}}\ and\ \bibinfo {author} {\bibfnamefont {B.}~\bibnamefont
  {Vinter}},\ }\bibfield  {title} {\bibinfo {title} {Many-body correlation
  effects on the (110) and (111) silicon inversion layers},\ }\href
  {https://doi.org/10.1103/PhysRevB.28.3639} {\bibfield  {journal} {\bibinfo
  {journal} {Phys. Rev. B}\ }\textbf {\bibinfo {volume} {28}},\ \bibinfo
  {pages} {3639} (\bibinfo {year} {1983})}\BibitemShut {NoStop}%
\bibitem [{\citenamefont {Bloss}\ \emph {et~al.}(1979)\citenamefont {Bloss},
  \citenamefont {Sham},\ and\ \citenamefont
  {Vinter}}]{blossInteractionInducedTransitionLow1979}%
  \BibitemOpen
  \bibfield  {author} {\bibinfo {author} {\bibfnamefont {W.~L.}\ \bibnamefont
  {Bloss}}, \bibinfo {author} {\bibfnamefont {L.~J.}\ \bibnamefont {Sham}},\
  and\ \bibinfo {author} {\bibfnamefont {V.}~\bibnamefont {Vinter}},\
  }\bibfield  {title} {\bibinfo {title} {Interaction-{{Induced Transition}} at
  {{Low Densities}} in {{Silicon Inversion Layer}}},\ }\href
  {https://doi.org/10.1103/PhysRevLett.43.1529} {\bibfield  {journal} {\bibinfo
   {journal} {Phys. Rev. Lett.}\ }\textbf {\bibinfo {volume} {43}},\ \bibinfo
  {pages} {1529} (\bibinfo {year} {1979})}\BibitemShut {NoStop}%
\bibitem [{\citenamefont {Cole}\ \emph {et~al.}(1981)\citenamefont {Cole},
  \citenamefont {McCombe}, \citenamefont {Quinn},\ and\ \citenamefont
  {Kalia}}]{coleEvidenceValleyOccupancyTransition1981}%
  \BibitemOpen
  \bibfield  {author} {\bibinfo {author} {\bibfnamefont {T.}~\bibnamefont
  {Cole}}, \bibinfo {author} {\bibfnamefont {B.~D.}\ \bibnamefont {McCombe}},
  \bibinfo {author} {\bibfnamefont {J.~J.}\ \bibnamefont {Quinn}},\ and\
  \bibinfo {author} {\bibfnamefont {R.~K.}\ \bibnamefont {Kalia}},\ }\bibfield
  {title} {\bibinfo {title} {Evidence for a {{Valley}}-{{Occupancy Transition}}
  in {{Si Inversion Layers}} at {{Low Electron Densities}}},\ }\href
  {https://doi.org/10.1103/PhysRevLett.46.1096} {\bibfield  {journal} {\bibinfo
   {journal} {Phys. Rev. Lett.}\ }\textbf {\bibinfo {volume} {46}},\ \bibinfo
  {pages} {1096} (\bibinfo {year} {1981})}\BibitemShut {NoStop}%
\bibitem [{\citenamefont {Tsui}\ and\ \citenamefont
  {Kaminsky}(1979)}]{tsuiObservationSixfoldValley1979}%
  \BibitemOpen
  \bibfield  {author} {\bibinfo {author} {\bibfnamefont {D.~C.}\ \bibnamefont
  {Tsui}}\ and\ \bibinfo {author} {\bibfnamefont {G.}~\bibnamefont
  {Kaminsky}},\ }\bibfield  {title} {\bibinfo {title} {Observation of {{Sixfold
  Valley Degeneracy}} in {{Electron Inversion Layers}} on {{Si}}(111)},\ }\href
  {https://doi.org/10.1103/PhysRevLett.42.595} {\bibfield  {journal} {\bibinfo
  {journal} {Phys. Rev. Lett.}\ }\textbf {\bibinfo {volume} {42}},\ \bibinfo
  {pages} {595} (\bibinfo {year} {1979})}\BibitemShut {NoStop}%
\bibitem [{\citenamefont {Hossain}\ \emph
  {et~al.}(2021{\natexlab{b}})\citenamefont {Hossain}, \citenamefont {Ma},
  \citenamefont {{Villegas-Rosales}}, \citenamefont {Chung}, \citenamefont
  {Pfeiffer}, \citenamefont {West}, \citenamefont {Baldwin},\ and\
  \citenamefont {Shayegan}}]{hossainSpontaneousValleyPolarization2021}%
  \BibitemOpen
  \bibfield  {author} {\bibinfo {author} {\bibfnamefont {M.~S.}\ \bibnamefont
  {Hossain}}, \bibinfo {author} {\bibfnamefont {M.~K.}\ \bibnamefont {Ma}},
  \bibinfo {author} {\bibfnamefont {K.~A.}\ \bibnamefont {{Villegas-Rosales}}},
  \bibinfo {author} {\bibfnamefont {Y.~J.}\ \bibnamefont {Chung}}, \bibinfo
  {author} {\bibfnamefont {L.~N.}\ \bibnamefont {Pfeiffer}}, \bibinfo {author}
  {\bibfnamefont {K.~W.}\ \bibnamefont {West}}, \bibinfo {author}
  {\bibfnamefont {K.~W.}\ \bibnamefont {Baldwin}},\ and\ \bibinfo {author}
  {\bibfnamefont {M.}~\bibnamefont {Shayegan}},\ }\bibfield  {title} {\bibinfo
  {title} {Spontaneous {{Valley Polarization}} of {{Itinerant Electrons}}},\
  }\href {https://doi.org/10.1103/PhysRevLett.127.116601} {\bibfield  {journal}
  {\bibinfo  {journal} {Phys. Rev. Lett.}\ }\textbf {\bibinfo {volume} {127}},\
  \bibinfo {pages} {116601} (\bibinfo {year} {2021}{\natexlab{b}})}\BibitemShut
  {NoStop}%
\bibitem [{\citenamefont {Zhang}\ and\ \citenamefont
  {Das~Sarma}(2005)}]{zhangDensitydependentSpinSusceptibility2005}%
  \BibitemOpen
  \bibfield  {author} {\bibinfo {author} {\bibfnamefont {Y.}~\bibnamefont
  {Zhang}}\ and\ \bibinfo {author} {\bibfnamefont {S.}~\bibnamefont
  {Das~Sarma}},\ }\bibfield  {title} {\bibinfo {title} {Density-dependent spin
  susceptibility and effective mass in interacting quasi-two-dimensional
  electron systems},\ }\href {https://doi.org/10.1103/PhysRevB.72.075308}
  {\bibfield  {journal} {\bibinfo  {journal} {Phys. Rev. B}\ }\textbf {\bibinfo
  {volume} {72}},\ \bibinfo {pages} {075308} (\bibinfo {year}
  {2005})}\BibitemShut {NoStop}%
\bibitem [{\citenamefont {Attaccalite}\ \emph {et~al.}(2002)\citenamefont
  {Attaccalite}, \citenamefont {Moroni}, \citenamefont {{Gori-Giorgi}},\ and\
  \citenamefont {Bachelet}}]{attaccaliteCorrelationEnergySpin2002}%
  \BibitemOpen
  \bibfield  {author} {\bibinfo {author} {\bibfnamefont {C.}~\bibnamefont
  {Attaccalite}}, \bibinfo {author} {\bibfnamefont {S.}~\bibnamefont {Moroni}},
  \bibinfo {author} {\bibfnamefont {P.}~\bibnamefont {{Gori-Giorgi}}},\ and\
  \bibinfo {author} {\bibfnamefont {G.~B.}\ \bibnamefont {Bachelet}},\
  }\bibfield  {title} {\bibinfo {title} {Correlation {{Energy}} and {{Spin
  Polarization}} in the {{2D Electron Gas}}},\ }\href
  {https://doi.org/10.1103/PhysRevLett.88.256601} {\bibfield  {journal}
  {\bibinfo  {journal} {Phys. Rev. Lett.}\ }\textbf {\bibinfo {volume} {88}},\
  \bibinfo {pages} {256601} (\bibinfo {year} {2002})}\BibitemShut {NoStop}%
\bibitem [{\citenamefont {Zhang}\ and\ \citenamefont
  {Sarma}(2006)}]{zhangCommentEffectsThickness2006}%
  \BibitemOpen
  \bibfield  {author} {\bibinfo {author} {\bibfnamefont {Y.}~\bibnamefont
  {Zhang}}\ and\ \bibinfo {author} {\bibfnamefont {S.~D.}\ \bibnamefont
  {Sarma}},\ }\bibfield  {title} {\bibinfo {title} {Comment on “{{Effects}}
  of {{Thickness}} on the {{Spin Susceptibility}} of the {{Two Dimensional
  Electron Gas}}”},\ }\href {https://doi.org/10.1103/PhysRevLett.97.039701}
  {\bibfield  {journal} {\bibinfo  {journal} {Phys. Rev. Lett.}\ }\textbf
  {\bibinfo {volume} {97}},\ \bibinfo {pages} {039701} (\bibinfo {year}
  {2006})}\BibitemShut {NoStop}%
\bibitem [{\citenamefont {Tanatar}\ and\ \citenamefont
  {Ceperley}(1989)}]{tanatarGroundStateTwodimensional1989}%
  \BibitemOpen
  \bibfield  {author} {\bibinfo {author} {\bibfnamefont {B.}~\bibnamefont
  {Tanatar}}\ and\ \bibinfo {author} {\bibfnamefont {D.~M.}\ \bibnamefont
  {Ceperley}},\ }\bibfield  {title} {\bibinfo {title} {Ground state of the
  two-dimensional electron gas},\ }\href
  {https://doi.org/10.1103/PhysRevB.39.5005} {\bibfield  {journal} {\bibinfo
  {journal} {Phys. Rev. B}\ }\textbf {\bibinfo {volume} {39}},\ \bibinfo
  {pages} {5005} (\bibinfo {year} {1989})}\BibitemShut {NoStop}%
\bibitem [{\citenamefont {Drummond}\ and\ \citenamefont
  {Needs}(2009)}]{drummondPhaseDiagramLowDensity2009}%
  \BibitemOpen
  \bibfield  {author} {\bibinfo {author} {\bibfnamefont {N.~D.}\ \bibnamefont
  {Drummond}}\ and\ \bibinfo {author} {\bibfnamefont {R.~J.}\ \bibnamefont
  {Needs}},\ }\bibfield  {title} {\bibinfo {title} {Phase {{Diagram}} of the
  {{Low}}-{{Density Two}}-{{Dimensional Homogeneous Electron Gas}}},\ }\href
  {https://doi.org/10.1103/PhysRevLett.102.126402} {\bibfield  {journal}
  {\bibinfo  {journal} {Phys. Rev. Lett.}\ }\textbf {\bibinfo {volume} {102}},\
  \bibinfo {pages} {126402} (\bibinfo {year} {2009})}\BibitemShut {NoStop}%
\bibitem [{\citenamefont {Das~Sarma}\ \emph {et~al.}(2015)\citenamefont
  {Das~Sarma}, \citenamefont {Hwang}, \citenamefont {Kodiyalam}, \citenamefont
  {Pfeiffer},\ and\ \citenamefont
  {West}}]{dassarmaTransportTwodimensionalModulationdoped2015}%
  \BibitemOpen
  \bibfield  {author} {\bibinfo {author} {\bibfnamefont {S.}~\bibnamefont
  {Das~Sarma}}, \bibinfo {author} {\bibfnamefont {E.~H.}\ \bibnamefont
  {Hwang}}, \bibinfo {author} {\bibfnamefont {S.}~\bibnamefont {Kodiyalam}},
  \bibinfo {author} {\bibfnamefont {L.~N.}\ \bibnamefont {Pfeiffer}},\ and\
  \bibinfo {author} {\bibfnamefont {K.~W.}\ \bibnamefont {West}},\ }\bibfield
  {title} {\bibinfo {title} {Transport in two-dimensional modulation-doped
  semiconductor structures},\ }\href
  {https://doi.org/10.1103/PhysRevB.91.205304} {\bibfield  {journal} {\bibinfo
  {journal} {Phys. Rev. B}\ }\textbf {\bibinfo {volume} {91}},\ \bibinfo
  {pages} {205304} (\bibinfo {year} {2015})}\BibitemShut {NoStop}%
\bibitem [{\citenamefont {Ahn}\ and\ \citenamefont
  {Das~Sarma}(2021)}]{ahnScreeningFriedelOscillations2021}%
  \BibitemOpen
  \bibfield  {author} {\bibinfo {author} {\bibfnamefont {S.}~\bibnamefont
  {Ahn}}\ and\ \bibinfo {author} {\bibfnamefont {S.}~\bibnamefont
  {Das~Sarma}},\ }\bibfield  {title} {\bibinfo {title} {Screening, {{Friedel}}
  oscillations, {{RKKY}} interaction, and {{Drude}} transport in anisotropic
  two-dimensional systems},\ }\href
  {https://doi.org/10.1103/PhysRevB.103.165303} {\bibfield  {journal} {\bibinfo
   {journal} {Phys. Rev. B}\ }\textbf {\bibinfo {volume} {103}},\ \bibinfo
  {pages} {165303} (\bibinfo {year} {2021})}\BibitemShut {NoStop}%
\bibitem [{\citenamefont
  {Stern}(1967)}]{sternPolarizabilityTwoDimensionalElectron1967}%
  \BibitemOpen
  \bibfield  {author} {\bibinfo {author} {\bibfnamefont {F.}~\bibnamefont
  {Stern}},\ }\bibfield  {title} {\bibinfo {title} {Polarizability of a
  {{Two}}-{{Dimensional Electron Gas}}},\ }\href
  {https://doi.org/10.1103/PhysRevLett.18.546} {\bibfield  {journal} {\bibinfo
  {journal} {Phys. Rev. Lett.}\ }\textbf {\bibinfo {volume} {18}},\ \bibinfo
  {pages} {546} (\bibinfo {year} {1967})}\BibitemShut {NoStop}%
\bibitem [{\citenamefont {Hwang}\ and\ \citenamefont
  {Das~Sarma}(2013)}]{hwangElectronicTransportTwodimensional2013}%
  \BibitemOpen
  \bibfield  {author} {\bibinfo {author} {\bibfnamefont {E.~H.}\ \bibnamefont
  {Hwang}}\ and\ \bibinfo {author} {\bibfnamefont {S.}~\bibnamefont
  {Das~Sarma}},\ }\bibfield  {title} {\bibinfo {title} {Electronic transport in
  two-dimensional {{Si:P}} $\delta$-doped layers},\ }\href
  {https://doi.org/10.1103/PhysRevB.87.125411} {\bibfield  {journal} {\bibinfo
  {journal} {Phys. Rev. B}\ }\textbf {\bibinfo {volume} {87}},\ \bibinfo
  {pages} {125411} (\bibinfo {year} {2013})}\BibitemShut {NoStop}%
\end{thebibliography}%

\end{document}